\documentclass[manuscript,screen,anonymous=false, balance]{acmart}

\usepackage{longtable}
\usepackage{tabularx}
\usepackage{enumitem}

\newcommand{\pogo}{\textit{Pok{\'e}mon GO}}
\newcommand{\post}{\textit{Pok{\'e}mon STAY}}

\newcommand{\revision}[1]{\textcolor{black}{#1}}
\newcommand{\revisionBegin}{\color{black}}
\newcommand{\revisionEnd}{\color{black}}
\AtBeginDocument{%
  \providecommand\BibTeX{{%
    \normalfont B\kern-0.5em{\scshape i\kern-0.25em b}\kern-0.8em\TeX}}}

\setcopyright{acmcopyright}
\copyrightyear{2022}
\acmYear{2022}

\acmJournal{TOCHI}

\acmBooktitle{}
\acmPrice{15.00}
\acmISBN{978-1-4503-XXXX-X/18/06}

\begin{document}

\title{\pogo{} to \post{}: How Covid-19 Affected \pogo{}  Players }

\author{John Dunham}
\affiliation{%
    \institution{Niantic x RIT Geo Games and Media Research Lab, Rochester Institute of Technology}
    \streetaddress{}
    \city{}
    \country{USA}}
    \email{jfd2017@rit.edu}
        
\author{Konstantinos Papangelis}
\affiliation{%
    \institution{Niantic x RIT Geo Games and Media Research Lab, Rochester Institute of Technology}
    \streetaddress{}
    \city{}
    \country{USA}}
    \email{kxpigm@rit.edu}

    
\author{Samuli Laato}
\affiliation{%
    \institution{University of Turku}
    \streetaddress{}
    \city{}
    \country{Finland}}
    
\author{Nicolas LaLone}
\affiliation{%
    \institution{University of Nebraska Omaha}
    \streetaddress{}
    \city{}
    \country{USA}}
    \email{nlalone@unomaha.edu}
       
\author{Jin Ha Lee}
\affiliation{%
    \institution{University of Washington, Seattle}
    \streetaddress{}
    \city{}
    \country{USA}}

\author{Michael Saker}
\affiliation{%
    \institution{City, University of London }
    \streetaddress{}
    \city{}
    \country{UK}}

\renewcommand{\shortauthors}{Dunham, et al.}

\begin{abstract}
Since its creation, the Location-Based Game (LBG), \pogo{}, has been embraced by a community of fans across the world.
Due to its recency, the impact of COVID-19 on the community of \pogo{} players is       underexplored.
We address how COVID-19 has impacted the players of \pogo{} by
building upon existing work focusing on player gratifications and impacts in \pogo{}.
\revisionBegin{}
Through semi-structured interviews, we provide a snapshot of the state of LBG play during unprecedented times.
These player testimonies demonstrate (1) the importance of in-person socialization to LBG,
(2) additional ways players use the game as a coping mechanism, and (3) how intentionality mediates player perceptions of people-place relationships.
In demonstrating these behaviors, we provide a glimpse of how a game that forces players to explore the world around them changed when going outside with friends became a source of danger.
\revisionEnd{}

\end{abstract}

\ccsdesc[500]{Human-centered computing~Human computer interaction (HCI)}

\ccsdesc[500]{Human-centered computing~Collaborative and social computing}

\keywords{Location-based Games, \pogo{}, COVID-19}

\maketitle

\section{Introduction}
Location-Based Games (LBG) are at the intersection of digital and physical spaces.
\revision{LBG are played on mobile devices, with digital assets overlaid on digital maps of physical environments.}
Importantly, for the players of these games, this imposition results in an augmented reality (AR) in that they adjust the meanings associated with physical spaces through a playful internal context.
LBGs have become a prominent aspect of the mobile gaming market in recent years.
\pogo{} \cite{PoGO}, for instance, has roughly 603 million players worldwide and has exceeded more than one billion dollars in sales in 2020 alone \cite{chapple_pokemon_2020,noauthor_global_2020}.
Concurrent to its record number of players, the COVID-19 global pandemic has dramatically affected how nearly everyone interacts with the space around them.
Stay-at-home orders, social distancing, and the transition to fully digital workplaces have had a marked effect on well-being and how people perceive the spaces around them \cite{ahmed_effectiveness_2018,bliss_your_2020,ettman_prevalence_2020}.

The COVID-19 pandemic has forced human cultures to rethink how they navigate human spaces \cite{ettman_prevalence_2020, prime_risk_2020}.
The restructuring of everyday life began with local, state, and national government bodies passing stay-at-home orders and mask-wearing to "flatten the curve" and not overwhelm supply chains and hospital systems.
For example, in the United States,  stay-at-home orders to slow the spread of COVID-19 in all but seven states, resulting in a near-universal shift in spatial reality for US residents \cite{moreland_timing_2020}.

While evidence supports the implementation of stay-at-home orders and social distancing to limit the spread of both the influenza virus \cite{ahmed_effectiveness_2018,jackson_effects_2014} and COVID-19 \cite{flaxman_estimating_2020,moreland_timing_2020}, the social, psychological, and economic effects of COVID-19 have forced some types of companies to reconsider their products entirely \cite{colley_geography_2017,walther_towards_2011}.
For example, LBGs are incredibly social experiences that rely on interacting with physical locations \cite{bhattacharya_group_2019, dunham_casual_2021}.
Prior to the pandemic LBGs encouraged in-person social gameplay \cite{kaczmarek_pikachu_2017,silva_digital_2009,vella_sense_2019}, required players to play in physical environments \cite{papangelis_performing_2020,wagner-greene_pokemon_2016,silva_playing_2008}, and had high mobility requirements \cite{ejsing-duun_location-based_2011}.
COVID-19 directly impacted these behaviors, forcing players to remain indoors.

The recontextualization of space and restricted mobility brought about by COVID-19 affords LBG researchers a unique opportunity to study the importance of mobility, space, and socialization in the LBG context.
\revision{As such, in the present research, we set out to explore (RQ1) how LBGs encourage mobility, (RQ2) the impact of external factors on people-place relationships, and (RQ3) whether socialization is the byproduct or the driving force behind LBG play.}

In exploring these concepts, we contribute a description of what these games offer players and how developers may tune LBGs to improve players’ physical and mental well-being not just in the pandemic but more broadly in more normal circumstances.
\revisionBegin{}
To this end, the present research presents the lived experiences of playing \pogo{} during the COVID-19 pandemic through thirty semi-structured exploratory interviews conducted via Zoom. The core contributions of this work are: (1) perceived changes of socialization, (2) coping mechanisms surrounding the game, and (3) changes to spatial relationships (See Table \ref{summary-of-themes} for a summary).

\begin{table}[!htbp]
\begin{tabular}{|l|l|}
\hline
\textbf{Theme}                                & \textbf{Description}                                                                                                                                                                                                                                                                                                    \\ \hline
COVID-19 Mediated Socialization Changes       & \begin{tabular}[c]{@{}l@{}}Socialization  migrated from chiefly in-person interaction to digital \\ interaction. To most, digital socialization didn't supplant in-person \\ interaction. Family groups, however,  showed resistance to \\ socialization changes, largely remaining in person.\end{tabular} \\ \hline
Coping Through \pogo{}          & \begin{tabular}[c]{@{}l@{}}\pogo{} was used by many of the participants as a means to \\ cope with stressors introduced by COVID-19. Typically, this was \\ through socialization, leaving the home, or acting as reminder to \\  pre-pandemic contexts.\end{tabular} \\ \hline
COVID-19 Redefined People-Place Relationships & \begin{tabular}[c]{@{}l@{}}Participants reported a shift in their perceptions of space and \\ place as a result of the pandemic. \pogo{} further redefined \\ these places to both ludic and pallatative contexts. \end{tabular} \\ \hline
\end{tabular}
\caption{Summary of Contributions}
\label{summary-of-themes}
\end{table}
\revisionEnd{}

This paper has the following structure: first, we present the relevant literature about LBG socialization and territoriality with an additional focus on the pandemic.
We then describe our methodology and outline how our data was gathered and analyzed.
After, we present the critical contributions are presented. 
We conclude the paper by outlining the following steps and implications while also discussing limitations in the research we conducted.

\section{Background}
This section first discusses socialization in online games and how it relates to LBGs.
We then move into an emergent phenomenon in LBGs, territoriality.
Finally, we will discuss COVID-19 and its impact on games in general and LBGs in specific.

\subsection{Socialization in \pogo{}}
Social interactions have long been an area of interest regarding traditional online games \cite{bartle_hearts_1996,de_kort_people_2008,nardi_strangers_2006}.
Online games, however, are not the only loci of social interaction in this context.
Through co-located play, games act as a meeting place for players to socialize to varying degrees \cite{voida_wii_2009}.
Some players socialize for socialization’s sake \cite{ito_hanging_2019}, while others use cooperative play to foster positive educational and health prospects \cite{takeuchi_new_2011}.
With LBGs, players engage in socialization patterns resembling MMOs and co-located socialization while also engaging in negotiated socialization.

LBGs augment the physical world with a layer of digital information that can be playfully interacted with \cite{benford_frame_2006}.
While goals are typically well defined, the rules of LBGs are generally more fluid.
Indeed, players will negotiate some rules in an LBG, while the game itself enables others \cite{silva_digital_2009}.
For example, while the game may define a point in physical space as a game location, the player defines what they consider to be allowable routes to it, considering personal circumstances, laws, and mores.
An LBG user must negotiate the boundaries and rules of the LBG for themselves at frequent intervals.
Player presence, game space possibility, and player intentionality are balanced with game goals to foster the rules experienced by the players  \cite{magerkurth_concepts_2007,walther_towards_2011,walther_towards_2011}.
Players, morals, mores, laws, and the game itself mediate the intersection of play intention and game rules.
Therefore, an LBG is an intrinsically social experience as the definition of the rules is rooted in elements of human socialization.

Studies of \pogo{}, the LBG studied in this work, have found high social interaction and communication in the players of the game \cite{ewell_catching_2020,militello_pokemon_2018}.
Further, \pogo{} also has positive associations with social \cite{ewell_catching_2020,vella_sense_2019}, mental \cite{watanabe_pokemon_2017}, and physical well-being \cite{althoff_influence_2016,kaczmarek_pikachu_2017,lindqvist_praise_2018}.
Critically, \pogo{} has been associated with potentially dangerous behaviors in distracted driving, biking, and walking, resulting in bodily harm \cite{lindqvist_praise_2018,wagner-greene_pokemon_2016}.

Next, \pogo{} frequently expresses socialization through player interactions with Points of Interest (POIs) in the form of Pokéstops and Gyms.
These POIs are static locations that supply players with the resources necessary to play the game representing locations ideal for co-located play.
In \pogo{}, a significant amount of play revolves around the Gym; it is a cooperative social event in which players battle powerful Pokémon to capture them \cite{noauthor_pokemon_2021}.
Attacks on gyms or battling a powerful Pokémon at a gym are called raids.
To participate in raids, the players must form groups to complete these challenges, as they are often too challenging to be accomplished by a single player.
Typically these groups are ephemeral and ad-hoc, lasting long enough to meet the challenge \cite{bhattacharya_group_2019,vella_sense_2019}. 
Groups can also be pre-formed in Discord, Facebook, or other game communities so they can engage in a series of raids together.
This behavior became more frequent with the introduction of “Raid Hour” (i.e., specific day/hour of the week when all gyms feature a specific raid boss) \cite{noauthor_pokemon_2021-1}.

When viewed through Arrow et al.’s theoretical lens of small groups \cite{arrow_small_2000}, these emergent groups collect contextual and local dynamics.
Contextual dynamics represent the region’s local culture, player availability, and unspoken and spoken rules around forming the groups.
In comparison, local dynamics represent the group’s activity and how they leverage tools and resources.
In an expression of local dynamics, it is not unheard of for temporary raiding groups to solidify into online communities which engage in co-operative raiding, collecting, battling, and other gameplay mechanics \cite{bhattacharya_group_2019}.

These communities lead to increased socialization, as participants would arrange in-person meetups facilitating social connectedness \cite{vella_sense_2019}.
In addition to raiding, Niantic (the developer of \pogo{}) frequently encourages social interaction through in-game events known as community days \cite{noauthor_pokemon_2021}.
These monthly events feature a specific Pokémon and provide gameplay bonuses and increased spawn rates, encouraging players to participate in areas with high densities of Pokéstops and Gyms.
These events would result in socialization above and beyond what is necessary to play the game, with players frequently organizing events in their local communities independent of Niantic and \pogo{} \cite{alavesa_unblurring_2020}.
The GO Fest, the biggest annual offline event, and other local events specific to particular regions offer similar opportunities for socialization \cite{noauthor_pokemon_2021-1}.

Socialization in \pogo{} is also frequently centered around the family, as many families play this game together \cite{saker_intergenerational_2021}.
Here, the familial playing of \pogo{} is indicative of Joint Media Engagement (JME) \cite{takeuchi_new_2011}.
Six conditions codify productive JME: mutual engagement, dialogic inquiry, co-creation, boundary-crossing, intention to develop, and focus on content, not control \cite{takeuchi_new_2011}.
As a corollary to this, Sobel et al. explored the dynamics of parent and child group players in \pogo{} \cite{sobel_it_2017}, demonstrating  \pogo{} expresses the six conditions of productive JME.
In this context, the game enabled parents to engage with their children: families could equally participate in challenges, dialogue about the game extended beyond gameplay, and families could learn about the game and their environment together \cite{saker_intergenerational_2021,sobel_it_2017}.
 
This study leverages existing literature, as it was conceived and conducted following the rise of COVID-19, to understand how the pandemic changed the social landscape of LBGs and \pogo{} more specifically.
The extant literature on sociality in \pogo{} is outlined in Table \ref{sociality-pogo}.
There exists a wide range of studies related to social interaction in \pogo{}, discussing both offline and online social gameplay \cite{laato_group_2021,rauti_learning_2020}.
Early on in 2016, shortly after the launch of \pogo{}, sociality did not seem to motivate players to play the game \cite{rauschnabel_adoption_2017}.
However, soon after raids were added sociality became an important part of locative play \cite{bhattacharya_group_2019,evans_motivations_2021}.
Alavesa and Xu \cite{alavesa_unblurring_2020} note that even before the COVID-19 pandemic forced people to quarantine themselves, a significant proportion of playing took place indoors.
The players integrated the game into their daily lives, as such, when they moved indoors so did their play \cite{alavesa_unblurring_2020,colley_geography_2017}.
At the same time, there is strong evidence that \pogo{} was able to activate players \cite{kato_can_2017,tateno_new_2016}, bring players to physical proximity to meaningfully cooperate with one another \cite{bhattacharya_group_2019}, increase their social connectedness \cite{vella_sense_2019} and to even influence the language and slang preferences of players \cite{laato_group_2021}.

\begin{longtable}[htpb]{|p{4.5cm}|p{8.5cm}|}
\hline
\textbf{Authors}& \textbf{Findings} \\
\hline
\endfirsthead
\multicolumn{2}{c}%
{\tablename\ \thetable\ -- \textit{Continued from previous page}} \\
\hline
\textbf{Authors}& \textbf{Findings} \\
\hline
\endhead
\hline \multicolumn{2}{r}{ \tablename\ \thetable\ -- \textit{Continued on next page}} \\
\endfoot
\endlastfoot
Alavesa and Xu, 2020 \cite{alavesa_unblurring_2020} & Playing is interlinked with players’ real life, including social life.AR images shared on social media suggest that indoor playing was commonplace among pre-pandemic players.\\ \hline
Bhattacharya et al., 2019 \cite{bhattacharya_group_2019} & Raids bring people together in temporal and spatial dimensions. 
The local group formation may solidify into online social groups.\\ \hline
Evans and Saker, 2019 \cite{evans_playeur_2019} & Playing \pogo{} led to people going out more and experiencing their environment differently. Equally, \pogo{} led to players playing in strange places - sometimes dangerous places - in a manner that differs from older locative games like Foursquare. Finally, this research spoke to players who primarily played for their children.\\ \hline
Evans et al., 2021 \cite{evans_motivations_2021} & Friendship maintenance and relationship initiation are strong motivators for \pogo{} players to play the game.\\ \hline
Laato et al., 2021 \cite{laato_group_2021} & LBGs are a mixture of online and offline interactions that are heavily interlinked.Players’ teams influence their linguistic identity and attitudes towards the game.\\ \hline
Rauschnabel et al., 2017 \cite{rauschnabel_adoption_2017}  & Among various uses and gratifications, sociality did not significantly lead to attitude towards playing \pogo{} nor intention to continue playing.However, social pressure led to the intention to continue playing and making in-app purchases.\\ \hline
Rauti et al., 2020 \cite{rauti_learning_2020} & \pogo{} brings players together in temporal and spatial dimensions through trading, lures, battles, and raids.In addition, the game provides asynchronous spatial, social activities such as gym battles and collecting gifts. The game also supports non-spatial social interaction.\\ \hline
Saker and Evans, 2021 \cite{saker_intergenerational_2021} & Playing \pogo{} saw families going out more and spending more time together.The familial playing of this game improved relations and allowed families to socialize around a joint activity. Yet, the social connections created were weak and transient. This speaks to the idea that family play of this game would not be impacted that much by the pandemic, as families often played alone. Finally, this research considered the impact of surveillance capitalism and how families were not concerned about personal data being used.\\ \hline
Tateno et al., 2016 \cite{tateno_new_2016} & \pogo{} may help with social withdrawal. The game is particularly useful for activating people who are otherwise disinterested in spending time (socially) outdoors.\\ \hline
Vella et al., 2019 \cite{vella_sense_2019} & Playing \pogo{} increases players’ sense of social connectedness.\\ \hline
\caption{Sociality in \pogo{} }
\label{sociality-pogo}

\end{longtable}

\subsection{Territoriality in LBGs}

Socialization in LBGs can also lead to antagonistic playful behavior due to players competing over their perceptions of the play space they inhabit.
This often manifests as expressions of territoriality through territorial claims \cite{papangelis_get_2017,papangelis_conquering_2017,silva_playing_2008}.
As players capture and recapture these spaces, they continuously redefine the space over time: renegotiating their relationship to the space they play.
Through these territoriality expressions, LBGs separate space from preconceived meanings and reassign them with new ones.
The recontextualizing of space through territoriality generally has social implications, a phenomenon highlighted in GeoMoments \cite{papangelis_performing_2020}.

Through the locative app GeoMoments, players assumed ownership over physical locations through a virtual layer.
App users were selective of the territory they controlled, with social factors frequently being the deciding factor of space’s value\cite{papangelis_performing_2020}.
To app users, a street corner in a nice part of town or shopping center carried with it the implicit social status of frequenting those places.
Through this locative app, players presented a performative self while recontextualizing the spaces they inhabited through the game’s lens.

To this end, LBGs alter a city’s legibility \cite{montgomery_making_1998} and imageability \cite{lynch_image_1960}, expanding the elements of the space to include not only physical paths and landmarks but also digital ones.
In a similar vein, \pogo{} expresses territoriality through the Gym POIs.
The game itself instructs players to align themselves with a team, an act of playful antagonism.
A team or player may capture and defend Gyms to assert dominance over a location.
In addition to the inherent rewards of digital ownership \cite{laato_primal_2020,woods_territoriality_2020}, players also receive benefits in the form of digital currency for maintaining control of the location \cite{noauthor_pokemon_2021-1}.

\subsection{COVID-19 and \pogo{}}

\begin{figure}[!htpb]
    \centering
    \includegraphics[width=0.2\textwidth]{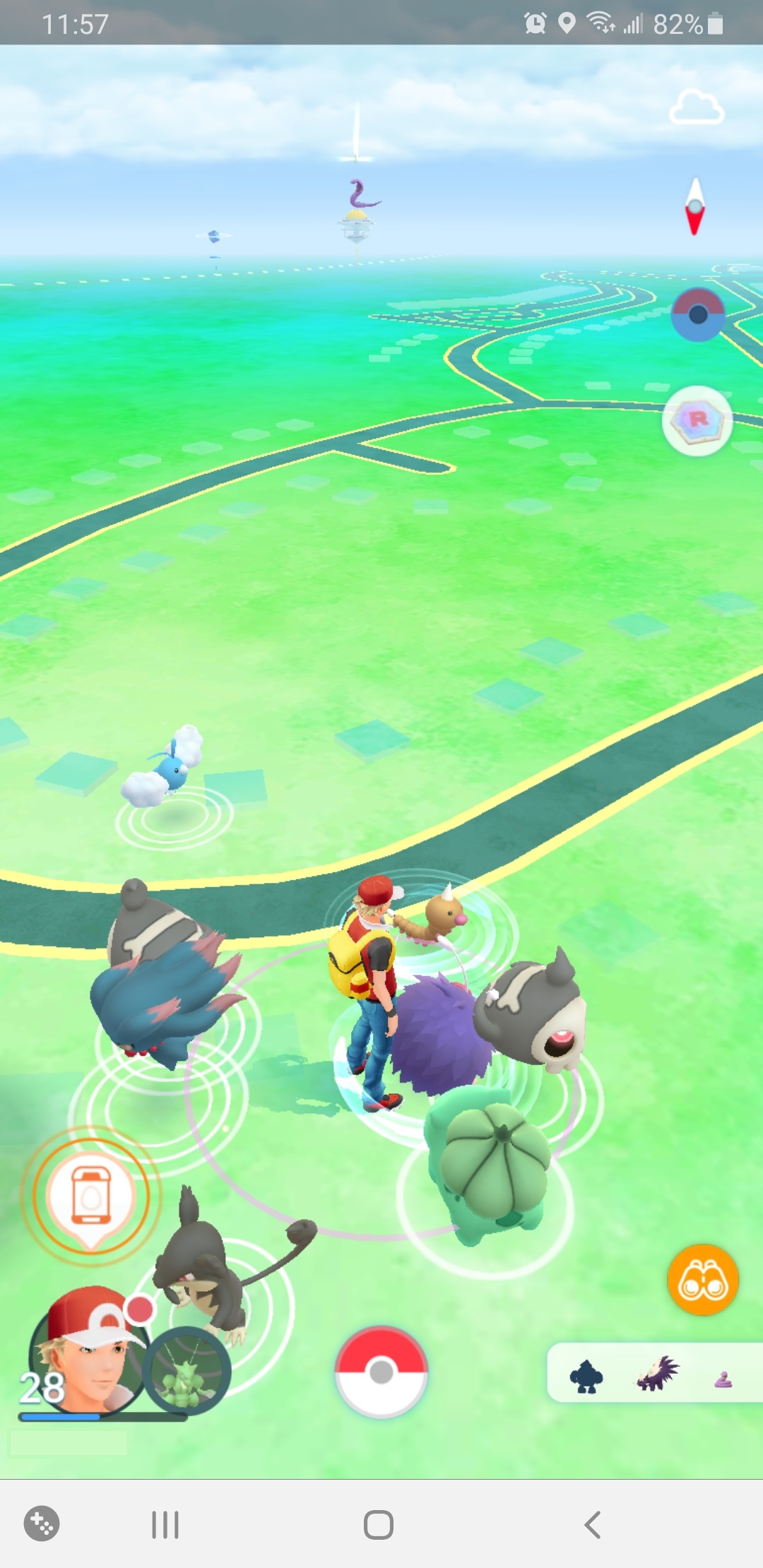}
    \caption{Main Interface of \pogo{} \cite{PoGO}, October 2020}
    \label{fig:pogo-screenshot}
\end{figure}

Since the start of the COVID-19 pandemic, nearly 82\% of global consumers have either played games or watched video game content \cite{noauthor_3_2020}.
In April 2020, the World Health Organization (WHO) partnered with game developers to promote the hashtag “\#PlayApartTogether” \cite{noauthor_games_2020}, encouraging socially distant gaming.
For locative media, such as \pogo{}, this was a potentially disastrous change due to gameplay’s highly social nature \cite{saker_intergenerational_2021}.
Yet, 2020 was a successful year for \pogo{}, generating over one billion US dollars in revenue in the first ten months alone \cite{chapple_pokemon_2020-1}.
Changes implemented by Niantic in 2020, therefore, may account for this success \cite{magerkurth_concepts_2007}.
However, the implications of these changes to socialization and territoriality are underexplored at present.

\begin{figure}[!htpb]
    \centering
    \includegraphics[width=.75\textwidth]{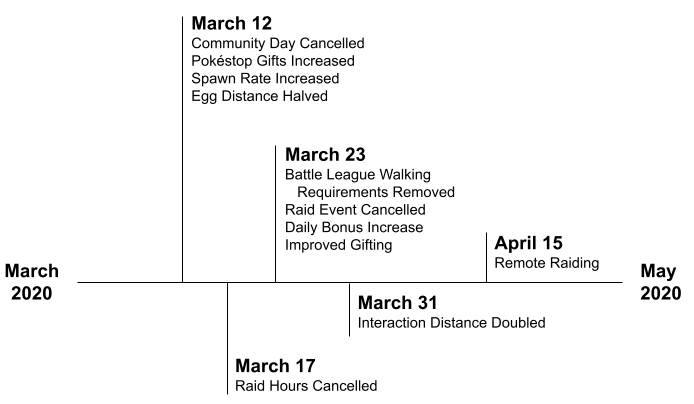}
    \caption{Timeline of changes in \pogo{} March 2020 - May 2020}
    \label{fig:pogo-timeline}
\end{figure}

To alleviate the impact of the COVID-19 pandemic, Niantic progressively rolled out changes during the 2020 calendar year to encourage continued play by its player base (See Figure \ref{fig:pogo-timeline}).
Niantic initially announced these changes on their official blog \cite{noauthor_pokemon_2021-1}.
Starting March 12, Niantic canceled the Abra community day in response to widespread lockdowns in the United States.
Niantic also halved Egg hatching distances (a mechanic which supplied players with Pokémon as a reward for walking), increased the gifts (a resource that gifted additional resources to peers) received from Pokéstops and increased the spawn rate for Pokémon \cite{noauthor_pokemon_2021-1}.

On March 17, the developers temporarily canceled raid hours.
This initial wave of changes diminished the critical social aspects of \pogo{} and reduced the general need to exit the home during the lockdown \cite{laato_location-based_2020}.
In March 2020, Niantic disabled the locative elements of the GO Battle League (an organized player versus player league) until May 1st.
Continuing the trend of disabling social aspects, Niantic canceled a planned raiding event on March 23 before indefinitely suspending raid hours.
On the same day, Niantic increased the daily bonuses, buffed gifts so players could give and receive more, and introduced a cheap rotating one Pokécoin (the game’s premium currency: ~ USD .01) bundle \cite{noauthor_pokemon_2021-1}.
The last significant change to \pogo{} came March 31st when the developers doubled the maximum distance at which players could interact with Gyms to increase accessibility in the pandemic’s opening weeks \cite{noauthor_pokemon_2021-1}.

Niantic’s changes have gradually chipped away at the game’s locative and social aspects, enabling players to play from home effectively \cite{saker_intergenerational_2021}.
Despite this, players still needed to play in the real world to really progress in the game.
Consequently, on April 15th \pogo{}’s introduced Remote Raid passes \cite{noauthor_pokemon_2021-1}.
Remote Raids represented a new way to raid in \pogo{}.

Here, players no longer needed to physically travel to a gym to challenge a raid boss, and they could even invite friends from across the world to virtually join in the challenge (for a nominal fee).
Further, Niantic removed the need to visit Pokéstops to acquire field research (quests) and gifts, making the game’s POIs and in-person social aspects entirely optional.
On May 24, Niantic even reintroduced Community Days branding them as “Play at Home Edition” \cite{noauthor_pokemon_2021-1}, cementing the transition from a game centered around locality and in-person socialization to one that encouraged static play and online social interaction.
    
Niantic has tweaked these gameplay changes in the intervening months to great success.
A sample of the main map screen after these changes were complete has been preserved in Figure \ref{fig:pogo-screenshot}.
In addition to being the best year on record for the game \cite{chapple_pokemon_2020-1}, the game continued to display positive aspects noted in earlier studies.
A survey by Ellis et al. found that players continued to use the game for socialization, exercise, and as an escape from the pressures of the pandemic \cite{ellis_covid-19_2020}.
Likewise, Laato et al. similarly found a continued drive by players to socialize during the pandemic \cite{laato_did_2020}, indicating that socialization remains a critical component of \pogo{}, regardless of the current circumstances.
In sum, then, our article seeks to understand how \pogo{} players’ adapted their approach to this game in response to the COVID-19 pandemic.

\section{Methodology}
\label{methods}
We conducted a series of semi-structured interviews with individuals who had experience playing \pogo{} during the pandemic.
Researchers gathered participants from several sources, including (1) Twitter and (2) Facebook posts, a (3) local discord server for \pogo{} players, and (4) the /r/pokemongo subreddit.
Participants were encouraged to sign up if they were willing to participate in a study exploring how the global pandemic and the associated changes to \pogo{} affected their gameplay. 
Initially, participants were not offered a reward for participating in the study; however, we later added an in-game reward valued at approximately \$10 to increase participation.
The first author retroactively provided the reward to participants interviewed before it had been added.

Registration for the study leveraged the online calendar tool Calendly \footnote{\url{https://calendly.com/}}.
We gathered basic demographic data and presented an informed consent document to participants before the interview with this tool.
Participants were required to provide age, pronouns, the number of years playing \pogo{}, and whether they perceived themselves as a “hardcore,” “casual,” or “other” player of the game.
Gender Identity was an optional parameter.
The informed consent document assured users that their data would be anonymous and confidential. 
Researchers allowed users to withdraw from the study at any point with no penalty.
This article provides pseudonyms to all named participants.

In total, 60 participants registered through this process, with 30 completing the interview and 1 participant giving a partial interview which was discarded in analysis (the interview had just covered demographic data before interruption).
The age of interviewed players ranged from 18 and 50.
The mean age was 30 years with a standard deviation of 8 years.
In terms of gender, 15 of the participants identified as male, 14 identified as female, and one declined to answer.
Most players considered themselves “gamers” (28/30) and had been playing \pogo{} for an average of 3 years with a standard deviation of 1.
\revisionBegin{}
Of the interviewed participants, interviewers confirmed 11 to be from the US, four from Europe, two from Africa, and the remaining from Canada (1), Asia (1), and South America (1) each.
The observed distribution of countries was consistent with Dunham et al.'s \cite{dunham_casual_2021} sampling of \pogo{} Reddit users, 10/30 of the participants didn't disclose their location.
For this reason, we eschew a more refined assessment of regionality for participant descriptions of local conditions.
\revisionEnd{}

Interviews began with a verification that the participants understood their rights in the informed consent document, confirmation of the supplied demographic information, and an additional question about gaming experience beyond \pogo{}.
The first author conducted the interviews in English through Zoom, recording discussions locally using Zoom’s record functionality.
Each interview took approximately 30 minutes.
Structurally the interviews were organized thematically, with the beginning of each interview asking the participants about their favorite feature of \pogo{}.
Participants were encouraged to mention in-game elements or social structures surrounding the game.

\revisionBegin{}
We analyzed our research questions in the context of the current literature to generate a set of related themes.
We identified five broad themes in this manner to structure our interview guide: (1) people-place relationships (including mobility), (2) socialization and social connectivity, (3) community engagement and culture, (4) outdoor activities (including exercise), and (5) motivations/gratifications.
As noted in Azungha \cite{azungah_qualitative_2018}, designing a qualitative interview guide in this manner allows us to ensure our research questions are addressed in each interview, as well as provide a base framework for us to structure our analysis of the interviews themselves.
The questions themselves were written as prompts with sub-prompts to give the interviewer control over how they elicited responses from our participants.
For example, one prompt was “Do you use \pogo{} as an excuse to leave home?” which had a potential sub-prompt of “Have you done this more or less since the onset of the pandemic?”

Broadly, our data analysis process leveraged the following structure: (1) data familiarization, (2) adjustment of the thematic framework and (3) coding of the data.
We transcribed the interviews with aid from the automated transcription product, Descript \footnote{\url{https://www.descript.com/}}, and then in cleaning the transcription manually, the first author enhanced their data familiarity.
As the first author conducted the data familiarization, they adjusted the thematic framework established in generating the interview guide.
While we removed no themes, additional themes, such as (1) family engagement and (2) \pogo{} as a coping mechanism, were added to the thematic framework as they had emerged from interviews.
Moreover, the core themes noted in the interview guide planning phase were adjusted to be umbrella themes for more granular concepts.
For example, sub-classifications of the community engagement theme were identified in this process, such as how remote raiding impacted the manner participants engaged in the \pogo{} community.
framework Therefore entering the coding phase there was already a base level thematic framework in place to structure our coding.
In this regard, we conducted our analysis initially deductively to ground further analysis in the extant themes of LBG.
This approach to thematic analysis borrows from the processes illustrated in adjacent work by Evans and Saker \cite{evans_playeur_2019}.

\revisionEnd{}



\section{Findings}


\revisionBegin{}
Our interviews paint a picture of a game changed deeply by external forces.
As Niantic re-engineered aspects of the game to reduce the impetus to travel to play (RQ1), participants indicated their relationships to the world around them and the greater \pogo{} community changed dramatically (RQ2).
Socialization saw dramatic shifts indicating not only did the game drive socialization but aspects of LBG inherently encouraged socialization, a fact sorely recognized by our participants (RQ3)
Unexpectedly, however, these interviews exposed the game as a source of comfort in our participants, allowing them to handle the myriad stressors of the pandemic better.
\revisionEnd{}

\subsection{Play Before and During the Pandemic}
Before the COVID-19 pandemic, socialization was a major part of playing \pogo{} for our participants (30/30).
Indeed, all participants described at least one offline social experience with \pogo{}.
Here, the sociability of these experiences manifests themselves in several ways: for some participants, it was simply a casual association with colleagues, while others recounted raids or recalled the excitement of taking part in community days.
These experiences varied both in frequency and intensity.

In addition to social play, many participants also played alone and derived pleasure from doing so.
When we asked which game experience they preferred, participants were roughly split down the middle, with nearly half focusing on social experiences (13/30), while others preferred solo experiences (17/30).
More precisely, solo players were interested in the “Gotta Catch ‘em All” experiences: completing their Pokédex, catching rare and powerful Pokémon, and taking on in-game challenges.
In contrast, social players focused on events involving other players, including raid hours and community days.

Moving forward, players who consistently played \pogo{} before the pandemic (25/30) recounted positive memories of doing so.
For example, one long-time player, Michelle [S005] (female 32), found it “fun to collect things and accomplish something, even if [she] couldn’t write anything” during her time in grad school.
For Michelle, the game effectively helped her relax during a difficult period in her life.
Likewise, Nicole [S016] (female 30) described the game as “a way of life” from the introduction of raids in 2017 again, underlining the positive impact of this experience.

During the pandemic, \pogo{} quickly adapted to a different kind of world, with the social experience of raids and community days shifting accordingly.
Here, health concerns, lockdowns, and in-game modifications led to many participants reporting significant changes to their playing experience.
In the following sections, we further explore these alterations and how they impacted participants’ experience of playing \pogo{}.

\subsubsection{Contextual Dynamics and Organizing Group Play}
Raiding is a major social activity of \pogo{}, in which players frequently form ad-hoc groups, as they are often too difficult to tackle alone.
Consequently, some participants (2/30) reported deliberately avoiding the more powerful raid bosses and only challenging targets they could combat in solo play.
For example, Alex [S019] (male 19) is a solo player who would explicitly target the “one-star [raids] that you can do on your own,” as opposed to the more lucrative legendary raids featuring more powerful Pokémon.

Equally, other players actively sought out people to engage in physical group play and raiding even during the pandemic.
In our study, 10/30 participants described some form of in-person raiding, which was largely formed on an ad-hoc basis.
Still, the pandemic impacted participants’ abilities to form groups.
For instance, Noel [P001] (female 23) recalled finding raiding partners before the pandemic was easy at her college.
As she explains, “people would gather together on campus to do raids … [, and] didn’t need much coordination … there would be like 10 or 20 people.”
Nonetheless, online coordination of raids saw a fair amount of occurrence before the pandemic, with 16/30 participants describing using Discord or Facebook to organize raiding.
Contextual dynamics of the region played a role in the frequency and prominence of online coordination.
Generally speaking, participants who reported ad-hoc play described experiences relating to colleges or city centers, \revision{indicating that density of players of the game impacted} how events would coalesce.

The participants suggest that raids were easy to engage with before the pandemic, which is, of course, noteworthy.
COVID-19 drastically changed this sentiment as people became less likely to leave the house and assemble.
\revisionBegin{}
While this was grounded in social distancing and quarantine orders in some cases, participants reported that fear of Covid-19 also restricted their movements.
One participant, Kyle [S024] (male 18), expressed they weren't "as comfortable because [they're] afraid of Covid and its long term effects,” limiting their movements in regards to the game.
Harold [S010] (male 18) mirrors this sentiment: "it's scary. You don't want to go out and get sick just for \pogo{}. It's not worth it."
\revisionEnd{}
Contextual dynamics, therefore, transformed raiding as ad-hoc raid groups became less common.

\revisionBegin{}
Similarly, the departure of students from college campuses exacerbated this situation in college towns, resulting in fewer players being available within those communities for communal play.
Tom [P005] (male 21) even cited this exodus as a reason to move to discord servers: "I started joining these Discord servers, ... [because] Covid ravaged our community. ... A lot of people left because [a college] is pretty much the worst place to be during the pandemic."  
Tom's experience is not unique as raids became almost exclusively organized online, indicating shifts in local dynamics.
\revisionEnd{}
Interestingly, despite players organizing more online, only 7/30 of the participants started using Discord, Whatsapp, or Facebook during the pandemic for this purpose, with the remaining participants continuing to use whatever platforms they had been using before.
Adopters of these networks were generally new players (4/30) or returning players to the game (8/30).


The perceived shift in communal styles, however, was generally received poorly by existing players, who did not appear to receive the same sense of gratification from the online platforms \revision{as they did from in-person interactions}.
For these participants, the rapid increase in the necessity of these networks left them feeling that organic connections within the game had declined.
As Brandon [S002] (male 30) explains: “you didn’t need Facebook [before the pandemic], you just met people organically, and now that’s a lot harder to do.”

\revisionBegin{}
While organic interaction was more challenging, it still happened with some frequency.
One participant who recently returned to the game, Jane [S026] (female 30), only became aware of the community during the pandemic when she recognized that "there were other people still claiming gyms, so there must be a community."
Jane's [S026] encounter with the \pogo{} community occurred within the core mechanics of the game, and while she didn't directly meet the players controlling the gym, she was able to recognize their existence.
Similarly, Connie [S007] (female 32), a veteran player of three years, only became aware of the \pogo{} community’s existence during the pandemic.
In Connie's [S007] case, it was a chance in-person encounter with a fellow \pogo{} player who invited her and her boyfriend to a group chat.
While initiated in physical space, the group chat made the existence of the community real to Connie [S007] as she realized  “it wasn’t just my boyfriend [who] plays [\pogo{}], but a whole lot of people.”
This appears to imply that organic community discovery is not exclusively rooted in physically tangible social interactions, as Connie's [S007] discovery was both physically tangible, but only became significant in a virtual context.

Regardless, some participants perceived a reduction in physically tangible social interactions, particularly in raiding, resulting in a blow to their enjoyment of the game.
\revisionEnd{}
Even with the help of social media platforms, several of the participants (5/30) stopped raiding at some point during the pandemic.
However, this situation changed when Niantic introduced the remote raiding feature.
Here, players could raid with other players irrespective of physical distances, with 26/30 of our participants engaging directly with this feature.

In many ways, this restructured the classification of dynamics and expanded their scope beyond previously studied manifestations.
In other words, a global community of players was now able to engage in local dynamics once exclusively in-person through digital platforms, and contextual dynamics of those groups became the particular platform’s culture.
Consequently, raids shifted from a locative experience to a digital one, resembling a more traditional massively multiplayer online (MMO) raid style (e.
g. World of Warcraft \cite{wow}).

As implicitly touched on above, many participants responded positively to the introduction of the remote raid system.
For Tom [P005] (male 21), it was a “week[ly] treat” that he would engage in, while for Kristen [S008] (female 30), a solo player, it felt like she was able to “virtually travel”.
In general, and on that note, solo players were happier with the remote raiding experience than more socially orientated players.
It gave these players a new opportunity to gather rare and strong Pokémon by joining raids listed on various social media platforms that involved minimal social obligations.

However, our study’s more socially oriented players did not find these changes entirely gratifying.
Nicole [S016] (female 30) noted that since the onset of the pandemic, she was not meeting new players during raiding and “not participating that much in a group [anymore].”
Remote raiding in Nicole’s case has supplanted in-person raiding and removed opportunities to meet new and old friends playing the game.
The game acted as a common ground for participants to communicate with people of disparate backgrounds, with conversations extending past the game’s strict confines.

Moreover, our participants reported that remote raids are frequently more challenging to organize than pre-pandemic in-person raiding.
Penny [S011] (female 44), a socially inclined player, found coordinating remote raids to be “exhausting sometimes.” For Penny, timing concerns, players having sufficient resources to raid, and the limited number of player invites made the experience difficult to manage.
Because of these organizational barriers to play, Penny had less time to focus on the social aspect of raiding, therefore, indicating that local dynamics suffered due to the changes brought on by the pandemic.

\subsubsection{Social Gratification and Extended Boundaries of Play}
For many socially orientated players, community days met with a similar fate as raids.
Before the pandemic, community days were social events where Pokémon would spawn more frequently, and rarer shiny versions would be easier to acquire.
These events would occur in a short, fixed period, encouraging players to carve out portions of their day to participate intentionally.
Following this, some participants would rearrange their work schedules to play during the event’s window.

For Jackie [S025] (female 34), for a “long time,” community days were the “biggest driver of the community at large.”
Players would hold raffles, share in-game resources known as lures to encourage more Pokémon, and generally socialize during the events.
However, with the advance of COVID-19, these events, which would organically appear in city parks, suddenly declined.
In part, and as one might imagine, this decrease was due to safety concerns over COVID-19.
Caring for high-risk family members, quarantine, social distancing orders, and general concern for one’s health were all stated as reasons for not participating in these events.
As our participant Sarah [S003] (female 50) notes, while it was not as “fun” as it had been previously, she could nonetheless engage in community days from “the couch”.

Participants witnessed changes in how community days functioned, with 11/30 participants directly referencing changes during their interviews.
Of the various changes noted, a reduction in the number of players playing at local hotspots for community days was most prominent.
Parks that once drew large crowds now drew smaller groups that would be less open to outsiders.
This change represented a shift in the culture surrounding community days commensurate with the contextual dynamics of raids following the pandemic.

The similarities extended to local dynamics as our participants reported that social media platforms became where the community aspects of community day would occur.
However, this transference of community to social media did not provide a social experience that was precisely comparable with what it had once been.
Indeed, none of our participants reported being fully gratified by the shift from the physical to the digital.
This particular finding indicates that the digital replacement of in-person interactions did not produce an experience that was as socially gratifying as it had been before COVID-19.

Interestingly, the extended duration of community days from 3 to 6 hours indirectly addressed players’ existing issues with the game.
\revisionBegin{}
Participants such as Jackie [S025] would frequently need to adjust their work schedules to participate in community days in pre-pandemic contexts, even going as far as "[requesting time] off ... or [getting] work shifts swapped .. just to do community days".
With the more lax duration, it was simpler for players to engage more passively, interweaving the event with their daily life.
\revisionEnd{}
For the most part, however, changes to the game were insufficient to overcome environmental factors to meet the observed goals of community day, as communal aspects declined during the pandemic.
In other words, while players could more readily participate in events, these changes reduced the sense of community observed by many participants, with 8/30 directly referencing this decline in their interviews.

\subsubsection{Family Play}
\label{family_play}
While a sense of greater community was reduced for the participants in our study during the pandemic, the more localized family playgroups thrived during the pandemic.
In our study, 11/30 players reported that they consistently played \pogo{} with family members.
\revisionBegin{}
Family play, therefore, appears not to be impeded by the pandemic forces that negatively impacted both raiding and community days.
Participants who played with family members not only continued to play the game, but the frequency of play increased.
Indeed, Families would use the game to spend time with their family, typically through some form of co-use.
\revisionEnd{}

This aspect of communal family play has not decreased during the pandemic - quite the opposite.
For instance, participant Mark [S009] (male 32) and his family played the game for the sake of their youngest member, his nephew (age 5).
Here, Mark [S009] and his family collectively played the game to provide some structure to his nephew’s lockdown and to “keep his mind occupied [so] he didn’t have to think about why we aren’t going out every day and stuff like that.”
For Mark [S009], the game was purposeful; it enabled him to connect with family members and help a loved one through a difficult time.

In general, familial locative play served to maintain and even strengthen communication in families of our study. 
Competitive families such as Penny’s [S011] (female 44) saw an increase in the expression of behaviors that had already codified their play: ”we've been [playing \pogo{}] all four years, but [the competition has] intensified over the quarantine.”
The game for Penny’s family relieved pandemic tensions and strengthened the bonds between family members as their primary form of entertainment.

As families typically quarantined together, there were more opportunities for in-person cooperative play than players who did not play with family members.
The majority of participants who reported playing with their families were happy with their experiences with the game during the pandemic.
Yet, there were notable exceptions where health concerns over COVID-19 were reported.
The family playgroup was more uniform in its positivity, with 10/11 reporting a neutral or better experience than the 14/19 participants who did not play with families.

Most of the participants who engaged in family play did not drastically change their play patterns with their family members.
Families who played together remotely with a messaging platform did so, while families who played in person continued to play in person.
Consequently, this kind of familial locative play was seemingly better able to withstand the changes brought about by the pandemic, as inversely demonstrated by the impact felt by other demographic groups interviewed.
We, therefore, suggest this resistance to pandemic forces might be symptomatic of participants not needing to alter their behaviors during periods of plays.

\subsection{Coping Through Locative Play}
Participant Michelle [S005] (female 32) recounted \pogo{}’s importance during graduate school before the pandemic.
To her, the game served as a “diversion” from graduate school’s inherent difficulties and stressors.
Several participants mirrored this sentiment during the pandemic, using the game to relieve stress from work, education, or even the pandemic itself.

Players generally appear to be continuing to use \pogo{} to enhance their mental health in the same ways as before the pandemic.
However, additional stressors from the pandemic have added new value to the game as a coping mechanism.
Some participants (7/30) joined or rejoined the game to escape from the realities of the pandemic.
\revisionBegin{}
For example, Harley [S018] (female 28), a returning player, was dealing with pandemic boredom when her brother-in-law suggested she "re-download \pogo{} [and she] said what the heck? You know I got nothing else better to do."
After rejoining the game, it became an excuse to engage in walks, providing a reason to leave home.
While Harley [S018] was using the game to escape boredom and encourage exercise, Mark [S009] (male 32) used the games with their family members to obfuscate the existential questions of the pandemic, as noted in section \ref{family_play}.
For Mark [S009], the game acted as a direct means to cope with the issues presented by the pandemic, and the mechanics afforded by \pogo{} allowed for a brief respite.

Harley [S018] and Mark [S009] typify participants matching this description: a family member or friend, introduces \pogo{} as something to do during the pandemic and the participant begins to play in earnest.
Interestingly, these participants universally described the game as an excuse to move and leave home.
Whether it be raiding, exploring the woods, or even hatching eggs, all of the participants who joined due to the pandemic leveraged the mobility aspects of LBG to great effect as a mechanism for coping.

However, for long-time players, the game was valuable for a different reason in dealing with pandemic stressors.
As was true in Michelle's [S005] case, these players would use the game to unwind after a long workday or take a break from coursework.
While exercise and familial play appear to have driven players who mentioned the pandemic in their reasons for returning to or joining the game, for veterans, the game served as a tether to times before the pandemic.
For example, Tom [P005] (male 21) had played the game for three years before being interviewed, meaning he had played for about two years in non-pandemic contexts.
For him, \pogo{} was “very relaxing and [took] [his] mind off [work]," which the pandemic had moved into his home.
The game allowed him "for at least ten minutes ... [to] feel like we're not in the middle of a horrible, disastrous pandemic."
Interestingly, Tom's [P005] relaxation was also linked to exercise and ambulation.

To Hank [S017] (male 33), the game served a similar purpose as a relief to pandemic woes, serving as an explicit escape.
While Hank [S017] was careful to note that "gaming in general [was an escape]," his story hinted at a more profound value in \pogo{} than traditional games.
Hank [S017] notes, "you [could] escape to a world where all [you] have to do is catch Pokémon ... and because [we're] a level 40 player's we don't have to worry about things running away. It's almost like a constant; if you open [the app] at home there's one or two spawn spots.
If you walk to the shop there are [more], it's comforting. ...
Despite the fact there is a virus that's raging around the world, this is constant."
To Hank [S017] \pogo{} allowed him to ground his reality to a familiar, consistent experience in a rapidly changing world.

Other participants, such as Brandon [S002] (male 30), also pointed directly at the LBG characteristics of \pogo{} as its value as a tool for relaxation.
For Brandon [S002] \pogo{} was a literal escape from his home during the pandemic, to him "no one wants to be stuck, cooped up at home ... [wouldn't] you rather be out exploring, being at least a little active?"
In this participant's case, his home transitioned to a space that captured him, but the game gave him an excuse to extricate himself from it.
Moreover, in playing, he had more opportunities to encounter green spaces, which he found to be a "kind of stress release" because it "[felt] great to be outside.”
While Brandon's [S002] reason to use the game to relax isn't necessarily tied to the pandemic, it has undoubtedly been exacerbated by it.

To Penny [S011] (female 44) the game was more holistically valuable to her health.
In addition to getting more exercise from the game, the game served as "mental exercising [through] mental stimulation" by giving her a "reason, purpose and place to go [beyond to just] walk around a grocery store or something."
For Penny [S011] \pogo{} gave her a purpose beyond just survival in the pandemic, injecting play into her life and engaging her mentally.
Penny's [S011] family also played the game with her, allowing the game to serve as a focal point of socialization with her family.
This socialization is largely playfully antagonistic with family members "setting up raids or showing off the newest shiny ... in group chat."
The competitive nature of play, coupled with necessary movement and adding purposeful objectives to her day, worked in tandem during the pandemic offering a respite from its challenges.

The exact shape and purpose of \pogo{} in the lives of our participants as a means of escape from the pandemic were non-standardized.
How intense local restrictions were, the health and well-being of family members, and even the type of work participants engaged in appeared to influence the manifestation of \pogo{} as a coping mechanism.
Moreover, exercise, green spaces, socialization, mental stimulation, and the game’s existence as a constant emerged as common themes in participant responses.
At its core, the game provided participants with structure in their lives to attach meaning important to them: a family-oriented person such as Mark [S009] rendered this structure as an apparatus facilitating family play.
For Brandon [S002], this structure acted as a reason to leave the house and explore.
Finally, Penny[S011] saw it as a combination of the two.
No matter how small or large the impact of \pogo{} on a participant’s means of dealing with the pandemic, it afforded them the capacity to redefine the space around them and the way they interacted with the world.
Most importantly, as their worlds shrunk under quarantine and social distancing \pogo{} allowed our participants to recontextualize their world not only as they saw fit but as they needed it.
\revisionEnd{}


\subsection{Spatial Alterations and Intentional Ambulation}

\pogo{} requires players to interact with physical space in a manner that is different from traditional games.
Before the pandemic, players left their homes to play the game, visiting in-game locations that serve as virtual landmarks used as waypoints to navigate and contextualize space.
One participant, Penny [S011] (female 44), “could probably tell you where every single PokéStop is in the city that [they] live in,” indicating a strong presence of in-game waypoints in her mental topography.
\revisionBegin{}
In this context, the PoIs act as an imagable object the players can use to build or restructure their mental maps of their surroundings.
The association of these digital locations with tangible in-game resources and out-of-game social rewards strongly highlights these locations to the game’s players.
\revisionEnd{}
The pandemic restructured the way people interact with the space surrounding them in general: commutes disappeared for many, places became taboo due to restrictions, and free movement became impossible in areas hit by quarantine measures.
\pogo{} was no exception, as players learned to restructure their interactions with the game and the spaces in which they played it during the pandemic, \revision{revealing some potential answers to RQ2 as people-place relationships saw noticeable changes}.


\revisionBegin{}
In one case, Michelle [S005] (female 32) related that "[she] used to go to the local university campus on community days to walk somewhere different, [but since the pandemic] it feels really wrong to go to the university campus."
To Michelle [S005], a place she had not only once played but also socialized with other players, had become a location restricted by unspoken social norms.
The campus was still accessible; however, the pandemic had contextualized the space and changed the game’s rules.

Michelle’s [S005] outdoor play was consistent with the remainder of our participants, with 27/30 describing leaving home to play the game.
That being said, pandemic forces had reduced the frequency, recontextualizing the outside world.
However, the consensus amongst these players was the game was still best played outside of the house, despite improvements allowing play from home.
In addition to recontextualizing space, how they navigated space to play changed as well as participants found themselves engaging in more frequent car play (10/30) or adjusting the routes (paths between game PoIs to maximize play) (10/30) they took when playing on foot.

Routes taken along clusters of PoI were described in 17/30 of the participants’ accounts as a typical play pattern during and before the pandemic.
This continued usage highlighted the importance of PoI in the game, with the pandemic exacerbating the value of PoI as players were unable to visit them to acquire the resources necessary to play the game as frequently.
Those who already had difficulty gathering resources (3/30) found the pandemic had been exacerbating them.

Regardless, all participants reported that they modified how they interacted with space regardless of pandemic fears as play styles shifted incidental to more intentional.
Before the pandemic, much of the play underpinning \pogo{} was incidental or ad hoc: visiting stops and catching Pokémon on the commute, in class, or while shopping.
Intentional play was not unheard of, particularly amongst the most dedicated and those who participated in raiding or pre-pandemic community days.
However, many participants (7/30) reported working from home during the pandemic, with all indicating experiences with shelter-in-place restrictions.

Naturally, this impacted our participants’ chances for incidental play as commutes, the locus of incidental play, had all but disappeared.
An unexpected consequence of the reduction in intentionality is a reduction in our participants’ spatial awareness (15/30) while playing the game.
As Noel [P001] (female 23), recalls "I [went] out to a new Lake and I found myself more engaged by the game because I was spinning these new stops rather than enjoying the Lake scenery. ... I put the game away, ... [and tried to] actually live in the moment and enjoy the scenery in front of me."
This reduction appears to be tied to the engagement models employed by \pogo{} (particularly spinning stops and catching Pokémon). 
However, reduction in spatial awareness appears to be largely agnostic to the pandemic, as all participants who described this phenomena recalled it applying in pre-pandemic contexts.
\revisionEnd{}

Still, a tension arises between participants being less aware of the space they inhabit and health concerns during the pandemic.
Several participants (6/30) indicated that their play was more cautious due to health and had become entirely intentional. 
They appeared acutely aware of others occupying the same space as them during the pandemic.
This change in perception may be in part because the pandemic has made intentional play “much more of a process,” as Tom [S004] (male 26) notes.
To play, the game players need to prepare themselves more rigorously: donning a mask and arming themselves with sanitizer before leaving the house.
As a result, physical artifacts exist on their person, acting as a constant reminder to be mindful of others.
While the locations fell into the background for our participants, the people entered the forefront.

While the physical locations fell into the background for the participants, the virtual ones were thrust more prominently to the forefront of their minds.
For players who could not travel as much, locations with PokéStops and Gyms suddenly became more valuable.
For example, Nicole’s [S016] (female 30) grocery store experience no longer simply revolved around a site to purchase essential resources and food.
Instead, the grocery store became a lifeline for continuing to play the game at a time when strict quarantine restrictions severely limited her ability to travel.
As travel saw a marked reduction for the participants, there were fewer opportunities to encounter PokéStops and Gyms incidentally.

\revisionBegin{}
Some participants had more significant restrictions preventing leaving their homes to play the game.
For some of these participants, such as Sarah [S003] (female 50), risking the lives of their at-risk family members for the sake of a game was not tenable.
The game became centered around the home to these participants, a pain point early in the pandemic as their resources and opportunities to play dried up.
However, as Niantic introduced new changes to the game, stationary play became more accessible, and participants in cities suddenly had access to PokéStops that were previously out of reach. 
For other players, gifts ensured they had a constant supply of Pokéballs.
Participant living rooms were suddenly an accessible place to play \pogo{}, or rather \post{}.
\revisionEnd{}

\section{Discussion}

\subsection{\revision{Socialization Redefined}}

The present research’s findings indicate that the socialization involved with playing \pogo{} shifted in response to the pandemic.
More precisely, the kind of socialization that was endemic of raiding and community days decreased following changes in society and the game due to COVID-19.
For the most part, these changes were brought about by the increased social distancing and isolation implemented to combat COVID-19.
Players lost the ability to interact personally with other players, or at the very least, saw a large decline in the feasibility of this form of interaction.

By ceasing to congregate in parks or around points of interest (POI) in the game, players enacted Silvia and Sutko’s assertion that, in part, players negotiate the rules of LBGs \cite{silva_digital_2009}.
That is to say, players changed the game’s rules, with social distancing and lockdowns acting as the driving force.
Players left the sidewalks and more actively utilized the platforms they had already leveraged for coordination.
In turn, Niantic shifted with the players adjusting the rules of their game, making the renegotiation of rules a two-way street \cite{silva_digital_2009}.
As noted in the background section, Niantic allowed players to thrive in more socially isolated contexts by changing the game’s core mechanics.

Yet, all players did not receive these changes positively, as highlighted in the shifting attitudes towards raiding.
Raiding underlines the interplay between player and developer action in negotiating \pogo{}’s rules, with lockdowns completely shutting down this aspect of the game.
In several cases, participants even noted that early raiding changes were primarily responsible for them playing less or individuals they played with stopping.
Some players directly called out the impact on their ability to socialize: it represented a loss of an integral aspect of the game to others.
Nonetheless, most of our players persisted through the raiding drought.

Only a third of the participants identified as hardcore players (very dedicated players) of \pogo{}, indicating that this may not be a factor vis-a-vis commitment to the game.
However, this may also be an artifact of players being bad at identifying their placement on the hardcore-casual continuum \cite{dunham_casual_2021}.
The participants generally underestimated their commitment level to the game, indicating that this effect was at play in the participants based on interviewer assessments.
Reconsidering our players as being largely more dedicated to the game than they reported recontextualizes these findings.
Under this interpretation, player dedication appears to be a reasonable predictor of whether this renegotiation of the rules was palatable to the players who considered raiding their favorite experience.
Players sufficiently dedicated to the game found the other elements’ sum to outweigh the gratification loss from their favorite mechanic changing.
One player who appeared to be devoted casually to the game stopped playing after raiding became a virtual event.
In other words, this in-person socialization was the driving force for their play resulting in them all but quitting the game, \revision{indicating the answer to RQ3 is largely mediated by the player's particular gratifications and commitment.}

\subsubsection{\revision{Raiding Contextualized to Digital Spaces}}

The changes implemented by Niantic led to further renegotiations of raiding rules, revolving around the inclusion of online socialization as a vital component in the form of raid passes \cite{noauthor_pokemon_2021-1}.
Local dynamics \cite{arrow_small_2000} shifted to organize online wholly in response, but this online organization was already the norm for many participants.
Organization moved from identifying a raid and setting a meetup time to wrangling raid participants to ensure all participants could participate.
In this way, while the participants of our study appreciated the functionality of remote raiding, it did not replace the socialization offered by pre-pandemic raiding \cite{bhattacharya_group_2019}.
The increased complexity was less satisfying than the more straightforward ad-hoc pick-up groups observed in the study of Bhattacharya et al. \cite{bhattacharya_group_2019}.
Similar trends are emphasized in prior work, which has highlighted explicit benefits afforded by LBGs concerning in-person interactions \cite{kaczmarek_pikachu_2017,silva_digital_2009}, and reinforced by participants who noted this discrepancy between in-person and online interactions.

The changes brought about by remote raiding extend beyond the discrepancy between in-person and online play.
Remote raiding created a new way to play \pogo{}, just as it reshaped a core element of the game.
Raiding already had a strong association with locative play in players’ minds, as in-person play forced players to move and get out of their homes.
In playing the game, players redefined the legibility \cite{montgomery_making_1998} of their surroundings and established connections between the game and the space they inhabited \cite{saker_intergenerational_2021}.

When raiding became remote, these prior associations of space may have influenced player perceptions, with multiple participants indicating that they felt as though they traveled while remote raiding.
The phenomenological sensation of travel felt by players may be an expression of the blurred nature of the boundaries of LBGs found in prior works \cite{bell_interweaving_2006,benford_frame_2006,gaver_ambiguity_2003}, as the virtual action and intent bleed into the player’s reality.
Through the ambiguity of the rules and the prior negotiated contracts of play, a stronger connection to physical space may have arisen through the game’s virtual space.
In this way, despite removing the need for strictly local play, the game at least partlymaintained some degree of locality.

\subsubsection{\revision{Engagement Through Socialization}}

While participants felt as though they were virtually traveling, their sense of community \cite{kim_probing_2020} was not as equipped to survive this digital leap.
Users of chat apps such as Discord, WhatsApp, or Facebook Messenger already had existing connections between socialization and these applications, so patterns of communication and behaviors were already established.
Players noted shifts in the contexts of the conversations held on these platforms concerning changes in the game, although it did not supplant pre-pandemic in-person interaction.

As social isolation took its toll on in-person interactions, digital platforms failed to rise to replace them.
Had the platforms been sufficient in replacing these interactions, players should not have experienced a reduction in their sense of community.
The decline of communal connectedness appears to be a manifestation of the “loneliness pandemic” \cite{luchetti_trajectory_2020,palgi_loneliness_2020}, indicating that even games like \pogo{} are not immune to this effect.

Participant response to changes in community days appears to support this assumption further.
Community days prior to COVID-19 were vibrant events that participants described asalmost fair-like.
Pandemic events were duller and did not provide the same gratification for our veteran participants, once again reinforcing the importance of in-person socialization.
However, participants who had not played the game before the pandemic did not feel this dissonance.
New players did not mourn the loss of in-person interactions, the interactions afforded by the game were fresher and provided them with relief from the pressures of the pandemic.

Moreover, players who engaged in family play saw nearly no decline in their sense of community or gratification with the game.
Prior work by Sobel et al. \cite{sobel_it_2017} alongside Saker and Evans \cite{saker_intergenerational_2021} on joint media engagement (JME) indicates that family play already had a substantial presence in LBGs such as \pogo{}.
It follows that despite the stressors imparted by the family by the pandemic \cite{prime_risk_2020}, participants who played with families appeared to find relief in \pogo{}.

To these players, their interactions with family members in the game introduced a sense of structure and normality to their lives.
Moreover, \pogo{} acted as a common goal or activity for the family to engage in, resonating with Saker and Evans \cite{saker_intergenerational_2021} study on families that played this game before COVID-19.
Indeed, this is also shown in Sobel et al.’s pre-pandemic work \cite{sobel_it_2017}, with the game serving as a focal point of engagement with their families.

In other words, the context of the game is less important than the social activities it enables, \revision{supporting socialization as a driving force of LBG play (RQ3)}.
Nonetheless, even with a reduced capacity during the pandemic, families could engage in playfully antagonistic competition (e.g., racing to catch better Pokémon or controlling gyms), collaborative play (e.g., raiding and joint usage of devices), and had an excuse to leave home.
As family players frequently lived together or at least were able to continue interacting in person, they were more insulated from the impacts of in-person interaction’s decline.
Through co-quarantining, players could continue to engage in locative play as a group, \revision{suggesting socialization acting as a primary mechanism by which LBGs encourage mobility in their players (RQ1)}.

In the larger group of players, co-located play declined among the participants.
The general structure of locative play saw a major shift in the participants, the most noteworthy of these changes being an increase in intentional play over incidental play.
Intentional play in \pogo{} has been observed in multiple contexts prior to the pandemic: most notably raiding \cite{aal_pokemon_2019,bhattacharya_group_2019}, community days \cite{aal_pokemon_2019,kim_it_2020}, and family play \cite{sobel_it_2017}.

\subsection{\revision{Coping Through Play}}

For the participants, incidental play accounted for much of their pre-pandemic play patterns, namely on the commute.
As noted in our Findings, a shift to work from home resulted in players no longer needing to commute, and previous opportunities to play disappeared.
Yet players still went out of their way to play the game, a finding mirrored by Laato et al.\cite{laato_did_2020}.
For some, this was to seek the game as a form of simple entertainment.
To others, the game was acting as a means of stress relief.
The participants reported that the game served as a moment of distraction from the stressors of the pandemic and even reminded them of a time before the pandemic.
This observation is supported by a recent study conducted by Ellis et al. \cite{ellis_covid-19_2020} indicating positive prospects for depression in players of the game.
Palliative qualities of \pogo{} have been reported before the pandemic in other studies \cite{watanabe_pokemon_2017} and by our participants.

\revisionBegin{}
How \pogo{} can have a palliative effect, however, is diverse.
To socially-minded players, the primary benefit of the game was its capacity to provide avenues of socialization.
Even if COVID-19 reduced in-person socialization, our participants frequently indicated the importance of socialization to their play and their gratification.
When considered in the context of the pandemic, known for its high rate of loneliness \cite{palgi_loneliness_2020}, the continued social outlet enabled by \pogo{} makes the game alluring to players.

In the case of family players, this benefit is further highlighted.
As families are frequently co-quarantined, in-person social interaction (the preferred interaction model of our participants) was still a reasonable interaction model.
Moreover, many family-oriented players had a pre-existing history of co-playing, allowing experiences in the game to be tied to pre-pandemic contexts, furthering the game’s capacity to reduce stress.
In other words, \pogo{} served as a means for families to engage in productive Joint Media Engagement (JME), as they cooperatively interact with media as a group \cite{sobel_it_2017,takeuchi_new_2011}.

Beyond the bounds of socialization, the game served as a distractionary activity.
While players engaged with the game, they could ignore the realities of the pandemic and instead worry about the routine presented by the game.
Whether maintaining control of their territory or simply catching new Pokémon the play elements represented an appreciated respite to our participants.
\revisionEnd{}
It stands to reason that players may have turned to the game as stressors increased, either consciously or not.
Moreover, the sensation of normality afforded to players by \pogo{} cannot be overstated.

Yet, within this normality, change emerged; as players visited locations purposefully to play the game, the contexts of those spaces changed for them.
For some, a supermarket transcended being only a place to get food; others redefined the meaning of local parks.
The participants described a mentality shift as locations attained new value based on their meaning in the game.
Pokéstops represented a lifeline to continued gameplay, indirectly relating them to improved mental health.
This recontextualization of space is supported by existing literature \cite{papangelis_performing_2020}, as the in-game contexts augmented or changed the real-world ones.

\subsection{\revision{Player Perceptions and the Intentionally of Play}}

As players’ meanings of locations change, so did their perceptions of the space around them.
Some players described the Pokéstop as the focal point of their play, as they were intentionally visiting these locations, often driving.
In this way, the gameplay was more driven by moving more intentionally from location to location, with the physical space being secondary to the virtual one.
Player intent appears to be critical in this regard, as players who did not report such reductions in awareness indicated that they deliberately engaged with their surroundings.

It appears that this deliberate awareness was in opposition to trends noted by Lindqvist et al. \cite{lindqvist_praise_2018}, wherein players were observed having difficulty engaging in their surroundings.
Interestingly, however, both before and during the pandemic, players appear aware of other individuals who occupy the same space.
Based on participant testimony, before the pandemic, players would routinely recognize other players and be aware of the reactions of non-players.
As LBGs are highly social experiences and blur the lines between play and reality \cite{walther_towards_2011}, this may be an extension of socialization in pre-pandemic interactions.

Pandemic recognition of others seems to be for a different reason, as health concerns were frequently noted when discussing others.
The physical artifacts carried by participants (e.g.mask, sanitizer) who noted increased awareness of others may have served as reminders to be more aware.
Additionally, as the physical space is unlikely to carry the same health threats as a potentially infected other, these spaces may take a back seat in the minds of the more health-conscious.
It is also noteworthy that players in self-described low restriction areas in our study did not typically have increased awareness of other people.

Changes in perceptions of space may also be the result of quarantine fatigue \cite{palgi_loneliness_2020}, as the home itself was recontextualized during the pandemic.
As a result of changes introduced by Niantic, the game could be more comfortably played from home.
A lesser need emerged for some players to leave home to play the game, particularly those with a Pokéstop nearby.
In this manner, \pogo{} shifted from being an LBG requiring offline interactions with people and space to effectively being a static solo or online experience.

Through this stark contrast, the home effectively became a point of interest in the LBG, contributing to the increase in play observed in many players and the rise in intentionality.
Players could now, from the comfort of their living room, perform nearly every action in the scope of the game.
Playing \pogo{}, then, was mechanically closer to playing a typical game than an LBG.

Nonetheless, tensions were still present for some players.
More precisely, the lack of having a Pokéstop nearby resulted in fewer resources.
Consequently, some players still needed to leave their homes to continue playing the game effectively, frequently engaging in fixed routes to play the game.
As detailed above, changes to commutes or routes chosen were not phenomena inspired by the pandemic.
Instead, routes were frequently embedded in player commutes and their navigation through their daily lives.

Here, participants noted strong memories for locations of PoIs and how they moved between them.
However, multiple participants reported changing routes, as locations became taboo for pandemic reasons: travel restrictions, shutdowns, and health concerns.
Once again players renegotiated the boundaries of play \cite{bell_interweaving_2006,benford_coping_2003,gaver_ambiguity_2003}, this time external factors pressuring the change.
Participants redefined space even further, once mundane paths to work became exotic as they began to understand their neighborhood more fully.
Likewise, new paths lost their luster as they were more frequently tread.

\revisionBegin{}
As these changes are all driven mainly by the pandemic, a potential answer to RQ2 emerges.
Sufficient disruptions in players’ lives impact how people-place relationships are mediated in the game.
What constitutes a sufficient disruption is linked to the player's pre-existing relationship.
However, external factors that reduce a player’s ability to navigate the space around them will invariably impact how the players engage in people-place relationships.
\revisionEnd{}

\subsection{\revision{Participants Want to Play}}

Ultimately, the participants just wanted to keep playing.
Our study identified the followings shifts in-game experience: (1) interaction with \pogo{} shifted from in-person to online, (2) a “soft reset” of sorts was experienced by the observed players, which redefined their relationships with real and virtual space around them, and (3) \pogo{} became increasingly meaningful beyond entertainment, with certain participants using this game to help them cope with the pandemic situation.

First, the combination of stressors caused by the pandemic (e.g., social distancing, quarantine, etc.) and the changes introduced by Niantic to the game (e.g., remote raiding) resulted in players shifting their interaction model with other players.
While generally, this shift did not fully replace the offline modalities, it produced new dynamics between players of the game and connected players who normally would not interact in the course of the game.
Moreover, some evidence suggests that transitioning a largely offline, locative game to a more statically defined online game may change how players interpret online interactions.
Users may make associations with the experiences of locative play to the non-locative, presenting users with the sensation of travel where there is none.

Second, restrictions to mobility reset a player's perception and relationship with the world around them beyond just the context of LBGs.
Participants reported changes to commutes, new patterns for leaving the house, and new perceptions of space’s value, including both within the home and in public spaces.
Much of this recontextualization resulted from the pandemic, local ordinances, and general cultural pressures.
However, even during the pandemic, \pogo{} influenced the people-place relationships experienced by the players.
More broadly, in terms of LBGs, real-world circumstances interact with the in-game content, forcing players to choose how they experience locations in terms of the game.

As the world shows signs of a potentially fragile and slow "new normal," the potential value of LBGs cannot be understated.
LBG researchers and developers are uniquely positioned to influence and enable the re-discovery of the city for players of such games.
Referring to previous work that has argued for the ability of LBGs to activate people \cite{kato_can_2017,tateno_new_2016}, shifts in gameplay have the potential to influence how people interact with cities.
Finally, the present research reinforces findings reported by both Laato et al. \cite{laato_did_2020} and Ellis et al. \cite{ellis_covid-19_2020} wherein \pogo{} was leveraged by players to cope with stressors of the pandemic in social and personal contexts.

\section{Conclusion}

The COVID-19 global pandemic represents not only a major disruption to daily life but an unprecedented upheaval to the regular operation of LBGs.
The present research presents the lived experiences of thirty LBG players to observe the perceived changes to the game from their point of view.
In pursuit of this goal, a series of semi-structured interviews were conducted which, amongst others, targeted key aspects of user experience: (1) the socialization around LBG, (2) the impact of family play, (3) player relationships to space, and (4) LBGs as a coping mechanism.


Socialization was recontextualized for many participants, with in-person interactions declining during the pandemic being supplanted by online ones.
As a corollary to this, it was also observed that online social media was not a replacement for the reduced in-person interaction.
However, new phenomena such as the sensation of virtual travel through the game manifested in response to the introduction of mobility agnostic gameplay mechanics.

Moreover, family units were found to be a reasonably static group in terms of perceived changes, with many continuing old behaviors or even increasing the amount they played during the pandemic.
Interestingly, these participants had more generally positive perceptions of the game during the pandemic than their cohorts who engaged in non-familial locative play.
Participants also observed changes in their routine in the LBG introduced by pressures of the pandemic.
These changes resulted in the recontextualization of the space around the participants, attaching new meaning and value to different locations in their lives.
Finally, the participants were found to use the game to psychologically cope with pandemic stressors.

\subsection{Limitations}
The limitations of this work are as follows.
First, the participants are predominantly recruited from online communities like Reddit’s /r/pokemongo board.
This sample may be more dedicated to \pogo{} than the general population, which, as noted in Dunham et al. \cite{dunham_casual_2021}, may result in the participants being disproportionately socially oriented.
This social orientation may have skewed our sample to have participants who perceived more stark changes to socialization than the average player.
Additionally, the participants are more likely to have been members of online communities by virtue of the participants being members of the online communities we advertised in.

Consequently, these participants may have a different perception of online communities.
Due to health concerns centered on the pandemic, interviews were exclusively conducted through Zoom.
Accordingly, in situ observations of play could not be conducted to reinforce participant interviews.
Similarly, researchers could not visit In-person raid hours and community days to confirm a reduction in participants.

\revisionBegin{}
The distribution of participant locations also limited this work, as we leveraged an opportunistic method to gather candidates. 
There were no guarantees regarding regional participant distributions.
Future work in this space may require a more precise accounting for player regionality to draw more specific conclusions.


Finally, given the temporal nature of LBGs and the pandemic’s unprecedented nature, this study represents a snapshot of a particular moment in the history of LBGs. As such conclusions must be considered from the state of the game when the present research was constructed and mediated through this lens in further studies. This work, however, opens several avenues for future work.
\revisionEnd{}

\subsection{Future Directions for LBG Research}
First, further research must be conducted into how players leverage \pogo{} and other LBGs to cope with stressors, particularly those relating to the pandemic.
Player relationships to space must also be investigated more closely.
A particular focus must be given to the notion of perceived travel through LBGs, as it represents an unexplored facet of LBGs.
Likewise, as the world reopens from the pandemic, the application of LBGs to sculpt the reopening of cities represents a novel area of research suggested by this work.
Additionally, as this work focused on urban play, future work could explore how the discovered phenomena play out in predominantly rural player communities.
 Ideally, this research should engage in ways that LBG can be ethically leveraged to allow users to rediscover and discover their cities through gameplay mechanics.
Researchers could further expand this work to compare and contrast urban and rural player communities in these contexts.

Additionally, while not new in the context of \pogo{} \cite{laato_location-based_2020}, the continued usage of the game as a means of coping with mental health concerns during the pandemic raises additional questions.
In what ways did this coping behavior manifest itself? Were there changes to user perceptions of play as a means of coping during the pandemic? How will the experience of the pandemic impact the relationship between the players and the game in a post-pandemic context? Our work indicates that there is evidence of using \pogo{} as a metaphoric crutch; however, interviews primarily focused on people-place relationships and socialization.

Future work should explore these questions and contextualize them to the broader LBG context.
This work should also investigate if and how LBGs can be directly targeted to better their player’s mental health.
As a final note, studies observing players’ return to in-person interaction should be conducted.
This work should serve as a companion to this work challenging the broader assertions made by this article discussing the core experiences of LBG play inferred from exceptional circumstances.

\bibliographystyle{ACM-Reference-Format}
\bibliography{main}

\begin{acks}
Niantic has funded this work through the Niantic X RIT Geo Games and Media Research Lab at the Rochester Institute of Technology.
\end{acks}

\appendix
\end{document}